\documentclass[12pt]{article}
\usepackage{hyperref}
\newcommand{\be}[3]{\begin{equation}  \label{#1#2#3}}
\newcommand{\bea}[3]{\begin{eqnarray}  \label{#1#2#3}}
\newcommand{\ee}{\end{equation}}
\newcommand{\eea}{\end{eqnarray}}
\newcommand{\ba}{\begin{array}}
\newcommand{\ea}{\end{array}}
\renewcommand{\arraystretch}{1.4}

\newcommand{\1}{{\mathbb 1}}

\let\Large=\large
\let\large=\normalsize

\usepackage{amssymb,bbold,euscript}

\begin{document}

\begin{flushright}
\hfill{UPR-1047-T}\\
\hfill{AEI-2003-062}\\
\hfill{hep-th/0308045}

\end{flushright}

\vspace{15pt}

\begin{center}{ \Large
{\bf  Supersymmetric Intersecting D6-branes and \\[2mm]
Fluxes  in Massive Type IIA String Theory
 }}

\vspace{20pt}

{Klaus Behrndt$^a$\footnote{E-mail: \tt behrndt@aei.mpg.de}
 and Mirjam Cveti\v c$^b $\footnote{E-mail:
{\tt cvetic@cvetic.hep.upenn.edu}. On sabbatic leave from the
University of Pennsylvania.}
 }
\vspace{15pt}

{\it $^a$ Max-Planck-Institut f\"ur Gravitationsphysik\\
Am M\"uhlenberg 1, 14476 Golm,  Germany\\[4mm]
$^b$ Institute for Advanced Study, Princeton, NJ 08540, USA }
\vspace{30pt}

\underline{ABSTRACT}
\end{center}

\noindent 
We study N=1 supersymmetric four-dimensional solutions of massive Type
IIA supergravity with intersecting D6-branes in the presence NS-NS
three-form fluxes. We derive N=1 supersymmetry conditions for the
D6-brane and flux configurations in an internal manifold $X_6$ and
derive the intrinsic torsion (or SU(3)-structure) related to the
fluxes.  In the absence of fluxes, N=1 supersymmetry implies that
D6-branes wrap supersymmetric three-cycles of $X_6$ that intersect at
angles of SU(3) rotations and the geometry is deformed by
SU(3)-structures. The presence of fluxes breaks the SU(3) structures
to SU(2) and the D6-branes intersect at angles of SU(2) rotations;
non-zero mass parameter corresponds to D8-branes which are orthogonal
to the common cycle of all D6-branes. The anomaly inflow indicates
that the gauge theory on intersecting (massive) D6-branes is not
chiral.

\newpage


\section{Introduction}

Constructions of four-dimensional N=1 supersymmetric solutions of Type
IIA string theory with intersecting D6-branes, wrapping three-cycles
of orbifolds (with an orientifold projection) provide an explicit
realization of supersymmetric ground sates of string theory with
massless chiral super-multiplets \cite{130,Bachas}. Explicit
supersymmetric solutions, based on $Z_2\times Z_2$ orbifolds, yielded
the first examples with the Standard-like model \cite{CSUI,CSUII,CPI}
as well as Grand unified (GUT) $SU(5)$ Georgi-Glashow model
\cite{CSUII,CPS}. [Non-supersymmetric intersecting D6-brane
constructions were initiated in \cite{new1,Blumetal,Ibanezetal,610} and
supersymmetric chiral constructions implicitly in \cite{AADS}.]  Such
constructions, when lifted on a circle to M-theory correspond to
compactifications of M-theory on compact, singular $G_2$-holonomy
metrics \cite{CSUII,W,AW,CSUIII}.

N=1 supersymmetry in D=4 imposes conditions on angles of the
three-cycles, wrapped by branes, relative to the orientifold plane. In
particular, for orbifold (toroidal) compactifications, the six-torus
$T^6$ can be written as a product of three two-tori $T^2$ and the
three-cycles as a product of three one-cycles (one in each $T^2$). In
this case the condition for supersymmetry \cite{130,CSUII} becomes a
condition that the sum of three angles, relative to the orientifold
plane, which have to sum to zero, i.e.  the rotation of the
three-cycles relative to the orientifold three-plane is an SU(3)
rotation. This condition is typically very constraining, since it has
to be satisfied for all three-cycles wrapped by various D6-branes;
if solved, it typically imposes conditions on the complex structure of
toroidal moduli \cite{CSUII}.

The consistency conditions \cite{Blumetal,CSUII} are equivalent
to the Gauss law for the D6-brane (positive charge) and O6-plane
(negative charge) sources; they ensure the charge cancellation
condition for D6-brane and O6-plane charges in the internal space and
impose constrains on the number of D6-branes and the wrapping numbers
of the supersymmetric three-cycles wrapped by D6-branes.

One can in principle generalize constructions on orbifolds (with
orientifold projection) to general Calabi-Yau manifolds $X_6$ (with
holomorphic $Z_2$ involution), see e.g.\ \cite{670} and refs.\
therein. The supersymmetry conditions on the three-cycles become
special Lagrangian conditions and the consistency conditions
reduce to an analog of the Gauss law for D6-brane and O6-plane sources
in the general Calabi-Yau background. The supersymmetry conditions can
be equivalently rephrased using SU(3)-structures generated by the
non-trivial fluxes. Note however, that unlike the orbifold
compactifications, where one has explicit conformal field theory
techniques to calculate the spectrum and couplings of the resulting
theory, for general Calabi-Yau compactifications techniques of
algebraic and differential geometry may not suffice to solve
explicitly consistency conditions and determine the full spectrum and
correlation functions.

The purpose of this paper is to address modifications of such
constructions due to the presence of fluxes associated with the
Ramond-Ramond (R-R) and Neveu-Schwarz-Neveu-Schwarz (NS-NS) closed
string sectors.  Our primary focus would be on quantification of
modified conditions that ensure N=1 supersymmetry in four-dimensions.
The presence of fluxes typically modifies Gauss law for D6-brane and
O6-brane sources. This modification, is due to the transgression
(Chern-Simons) terms which act as a source on the right hand side
(rhs) of the equation for the D6-brane field strengths (see, e.g.,
\cite{CLP}). These transgression terms give a positive contribution
to the ``total'' charge and thus fluxes typically modify the charge
cancellation conditions by reducing the number D6-brane configurations.

A framework where we can study the effects of fluxes\footnote{There is
a growing literature on the subject of string compactifications with
fluxes which was initiated in \cite{Strominger,PS}, a partial list of
subsequent works includes, e.g., \cite{GVW,DRS,TV,600,260}, and for
recent work, quantifying effects in terms of deformations of the
original manifold (G-structures), see
\cite{CGLP,Gauntlettetal,350,Louisetal,LustetalII,650,660} and
references therein.}  for the intersecting D6-brane probes in a
straightforward way turns out to be within masssive Type IIA
supergravity \cite{110}; it contains the Chern-Simons term that
couples the D6-brane potential $C^{(7)}$ to the zero-form (mass) $m$
and NS-NS 3-form field-strength $H^{(3)}$:
\begin{equation}
  L_{CS}=\int_{M_4\times X_6} m\, H^{(3)}\wedge C^{(7)}\, .
\label{CS}
\end{equation}

Note, that while massive Type IIA supergravity provides a
straightforward framework that via the Chern-Simons term (\ref{CS})
couples D6-brane probes to supergravity NS-NS three-form fluxes, there
are other possibilities. Within massless Type IIA supergravity with
off-diagonal metric components on $X_6$ turned on, the kinetic energy
term for the R-R sector fluxes may induce an effective transgression
term that couples R-R sector and metric fluxes to $C^{(7)}$ \cite{CU}.
Such examples could be related by T-duality to Type IIB configurations
with D3-brane probes and self-dual 3-form fluxes as studied in
\cite{GKP,KachruetalI,new2, KachruetalII} and for generalizations to
magnetized D-branes, see \cite{BLT,CU}.  In this paper we would like
to capture the explicit structure of the intersecting D6-branes in the
presence of NS-NS fluxes within massive Type IIA superstring theory.

The Chern-Simons term (\ref{CS}) turns out to modify the equation of
motion for $C^{(7)}$ as:
\begin{equation}
d \, F^{(2)}=m\, H^{(3)}
\label{trans}
\end{equation}
where the two-form $F^{(2)}$ is the magnetic field strength of
$C^{(7)}$.  Eq. (\ref{trans}) provides a modification of the
original consistency conditions on the number of D6-branes,
wrapping specific three-cycles, which are now modified by the rhs
of (\ref{trans}). While for orbifold backgrounds the charge
cancellation condition can be solved explicitly \footnote{Note
however that the quantization conditions for fluxes on orbifolds
can be subtle, c.f., \cite{BLT,CU}, one can over-saturate the
charge cancellation leading to the introduction of anti-branes
which explicitly break supersymmetry of the configuration.}, for a
general Calabi-Yau compactification (with $Z_2$ involution) these
conditions are complicated and the explicit solutions may be hard
to find.

Our main focus, however, will be on the modifications of the
supersymmetry conditions and classification of the internal torsion of
the resulting internal manifold which ensures N=1 supersymmetry in
four-dimensions.  We choose to turn on only the D6-brane sources
($C^{(7)}$), the mass parameter $m$ and the NS-NS three-form
$H^{(3)}$.  Within this framework, our approach shall be general;
we shall neither impose a priori conditions on the structure of
the internal manifold nor shall we impose a priori conditions on the
D6-brane and NS-NS three-form flux configurations. Such conditions
will be derived as a consequence of supersymmetry conditions,
i.e. from the Killing spinor equations of massive Type IIA
supergravity. (This is analogous to the analysis in Type IIB string
theory with D3-branes in the presence of three-form R-R and NS-NS fluxes
\cite{150}.)

The upshot of the analysis yields strong constraints on the
allowed D6-brane configurations: in the presence of mass parameter
$m$ and NS-NS fluxes D6-branes should intersect only at angles
compatible with SU(2) (and not SU(3)) rotations, in order to
preserve N=1 supersymmetry in four-dimensions, i.e. the
G-strucuture of the internal manifold is that of SU(2). Without
NS-NS fluxes (and $m$=0) the massless spectrum at the intersection
of D6-brane probes (rotated by SU(2) angles) corresponds to those
of N=2 hypermultiplets and is therefore {\it non-chiral}. However,
in the presence of fluxes the supersymmetry is broken down to N=1
and we expect that the spectrum changes. By studying anomaly
inflow we shall also address   whether the gauge-theory on the
world-volume of the D6-branes is chiral.

As a warm-up and to elucidate the derivation of the supersymmetry
conditions (and the intrinsic torsion of underlying compact
manifold) we shall also explicitly address supersymmetry
conditions for intersecting D6-brane probes without fluxes, thus
reducing the analysis to the framework of massless IIA superstring
theory. In this case we obtain SU(3) torsion classes for the
resulting internal space, which of course have a natural lift on a
circle to M-theory on $G_2$ holonomy manifold. This derivation is
closely related to the studies in \cite{350}.

The paper is organized in the following way: In Section II we
define the supersymmetry transformations and the Ansatz for the
metric and fluxes. In Section III we decompose the 10-d spinors
into 4- and 6-d spinors and we use the standard technique to
define a fundamental two-form and a three-form for the
six-dimensional internal manifold $X_6$.  We distinguish between
two cases: in {\it case (i)} we assume that the 6-manifold has
only a single (chiral) spinor and in {\it case (ii)} we consider
two 6-d spinors.  The existence of two spinors is equivalent to
the existence of a nowhere vanishing vector field $v$ on $X_6$ and
this vector breaks the SU(3) invariance of the supersymmetry
projectors in {\it case (i)} to SU(2) invariance in {\it case
(ii)}. In Section IV we explicitly derive the supersymmetry
conditions: {\it case (i)} is appropriate to the massless case,
i.e.\ a D6-brane background without fluxes, whereas {\it case
(ii)} corresponds to the massive case which, in the absence of
4-form fluxes, requires the existence of a vector field on $X_6$.
In the limit $m=0$, the NS-NS-three form flux is also turned off
and both cases coincide. In Section V we discuss the back reaction
on the geometry.  The flux deformations correspond to specific
SU(3) structures, i.e.\ the internal space is Calabi-Yau with
torsion and we show that two of the five torsion components are
zero. In fact, the vector field $v$ implies that the SU(3)
structures are broken to SU(2) structures.  In Section VI we
address the anomaly inflow  in the presence of NS-NS 3-fluxes as
well as  D6-brane, NS5-brane and D8-brane sources. We conclude
that there is no anomaly inflow, thus indicating that the D6-brane
world-volume  gauge theory is not chiral. In Section VII conclude
with a discussion of a number of open questions and possible
generalizations of our approach. In particular,  we  comment on
the structure of the four-dimensional superpotential and point out
possible generalizations, by  adding  R-R 4-form fluxes in the
massive Type IIA case.


\section{Supersymmetry variations}


In massive type IIA supergravity, one introduces a gauge invariant
2-form by
\[
F^{(2)} = m B + d A^{(1)}
\]
where the R-R 1-form $A^{(1)}$ can be gauged away to give a mass to
the NS-$B$-field. In the field equation for the NS-NS
3-form\footnote{In the rest of the paper we suppress the superscript
$(3)$ on $H^{(3)}$ and use simply the notation $H$ and $F$ for the
NS-NS 3-form field strength and 2-form potential, respectively.}
$H=dB$, the 2-form $m \, F^{(2)}$ appears as a source term and hence,
whenever $m\, F^{(2)}\neq 0 $, also the NS-NS 3-form has to be
non-zero and has to be included into our consideration.

We will especially be interested in the modification of the
supersymmetry constraints on intersecting D6-brane configurations due
to a non-vanishing mass parameter. Hence, in the limit $m=0$ we get
back to the standard D6-brane configuration which couples to the
one-form potential $A^{(1)}$ only. The massive $B$-field also enters
the 4-form field strength and in order to ensure the absence of
$F^{(4)}$:
\[
F^{(4)} = dC^{(3)} + 6 m B \wedge B = 0
\]
we have to impose the constraint (besides $C^{(3)}$ =0)
\be372
m\, B \wedge B = 0 \ .
\ee
For $m \neq 0$, one can gauge away the 1-form $A^{(1)}$ and this
constraint is equivalent to
\[
F \wedge F = 0 
\]
and means that we neglect effects due to 4-form fluxes and/or
D4-branes, which might be interesting in it own (Such configurations
may be related to the chiral brane-box model of \cite{HZ},
supplemented with additional supergravity fluxes. We briefly comment
on that in the Discussion section.)

Supersymmetry for purely bosonic background requires the vanishing of
the gravitino and dilatino supersymmetry variation and is equivalent
to the existence of a Killing spinor. In the canonical Einstein frame,
these variations are given in \cite{110}, but we are going to use the
string frame. Using the identity
\be030
\Gamma_M \Gamma^{N_1 \cdots N_n } = \Gamma_M^{\ N_1 \cdots N_n} + n
\delta_{M}^{\ [N_1} \Gamma^{N_2 \cdots N_n]} \ ,
\ee
the variations can be written as
\be040
\ba{rcl}
\delta \psi_M &=& \Big\{D_M  - {1 \over 4} H_M \Gamma_{11}
 - {1 \over 8} \, e^{\phi} \Big[ \, m    \Gamma_M  +
 ( \Gamma_{M} F  - 4 F_M ) \, \Gamma_{11}  \Big] \Big\} \epsilon \ ,
\\
\delta \lambda &=&  \Big\{ -{1 \over 2} \, \partial \phi
        -  e^{\phi}\Big[ {5 \over 8} m - {3 \over 8} F \, \Gamma_{11}
      - {1 \over 12} \, H \, \Gamma_{11} \Big] \Big\}  \epsilon
\ea
\ee
where we used the abbreviations
\be050
\partial \equiv \Gamma^M \partial_M \  , \quad
H = H_{PQR} \Gamma^{PQR} \  , \quad  H_M = H_{MPQ} \Gamma^{PQ} \ ,
\quad {\rm etc.}
\ee
Since we are interested in a compactification to a flat 4-d Minkowski
space, i.e. up to warping $Y_{10}={ R}^{(1,3)}\times X_6$, we write
the metric Ansatz as
\be060 ds^2 = e^{-2U(y)} \Big[-dt^2 + d\vec x^2  + h_{mn}(y) dy^m
dy^n \Big] \ .
\ee
Consistent with this Ansatz is the assumption that the fluxes
associated with the  2-form $F$ (and 3-form  $H$)  have non-zero
components only in the internal space $X_6$:
\be062
F = F_{mn} dy^m \wedge dy^n \quad , \qquad
H= H_{mnp} dy^m \wedge dy^n \wedge dy^p \ .
\ee
The $\Gamma$-matrices are decomposed as usual
\be070
\ba{l}
\Gamma^{\mu} =  \hat \gamma^{\mu} \otimes {\mathbb 1}
\quad, \qquad
\Gamma^{m+3} = \hat \gamma^5 \otimes \gamma^m
\quad , \qquad
\Gamma^{11} = - \hat \gamma^5 \otimes \gamma^7 \ , \\
\hat \gamma^5 = i \hat \gamma^0 \hat \gamma^1 \hat \gamma^2 \hat \gamma^3
    \quad ,\qquad
\gamma^7 = i \gamma^1 \gamma^2 \gamma^3 \gamma^4 \gamma^5 \gamma^6
\ea \ee
and we use the Majorana representation so that $\Gamma^{11}$,
$\hat \gamma^\mu$ are real and $\hat \gamma^5$, $\gamma^7$ and
$\gamma^m$ are imaginary and anti-symmetric.

With these expressions, it is now straightforward to decompose
the supersymmetry variations into external an internal components.
With our metric Ansatz the covariant derivatives can be written as
\[
\ba{l}
D_\mu = -  {1 \over 2} \Gamma_\mu^{\ m} \partial_m U = -
{1 \over 2} \hat \gamma_\mu \hat \gamma^5 \otimes \partial U \ ,
\\
D_m = \nabla_m - {1 \over 2} \Gamma_m^{\ n} \partial_n U
=  {\mathbb 1} \otimes [ \nabla_m - {1 \over 2} \gamma_m^{\ n} \partial_n U]
\ea
\]
where $\partial \equiv \gamma^m \partial_m$ and $\nabla_m$ is the
covariant derivative with respect to the metric $h_{mn}$.  Thus, the
external components of the gravitino variation become
\[
\delta \psi_\mu =  -{1 \over 2} \hat \gamma_\mu \otimes {\mathbb 1}
     \Big[e^U \hat \gamma^5 \otimes \partial U +
       {1 \over 4} e^{\phi} \, \Big( m
      [ {\mathbb 1}\otimes {\mathbb 1} ] -
       e^{2U} \hat \gamma^5 \otimes F \gamma^7 \Big)  \Big] \epsilon = 0 \ ,
\]
($F \equiv F_{mn} \gamma^{mn}$) and it is solved if
\be090
 e^U (\hat \gamma^5 \otimes \partial U ) \, \epsilon
 = -{1 \over 4} e^{\phi} \, \Big( m
      [ {\mathbb 1}\otimes {\mathbb 1} ] -
       e^{2U} \hat \gamma^5 \otimes F \gamma^7 \Big)  \epsilon \ .
\ee
Using this expression, we can now bring the internal components of the
gravitino variation into the form
\be121
\delta \psi_m =  \Big[ {\mathbb 1} \otimes \Big( \nabla_m +
{1 \over 2} e^{\phi + {U\over 2} } F_{m}  \gamma^7 \Big) -
{1 \over 4}\, \hat \gamma^5 \otimes  e^{{3 \over 2}U} H_m \gamma^7 \Big] \hat
\epsilon \quad , \qquad \hat \epsilon \equiv e^{U \over 2}
\epsilon
\ee
with $F_m \equiv F_{mn} \gamma^n$, $H_m \equiv H_{mpq} \gamma^{pq}$.
In a similar way, we can also simplify the dilatino variation
and find
\be721
\delta \lambda = - {1 \over 2} \Big[ \hat \gamma^5 \otimes
e^U \, (\partial\phi + 3 \partial U) +{\mathbb 1} \otimes
e^{\phi} \Big( m + {1 \over 12} e^{3U}\, H \gamma^7 \Big) \Big] \epsilon \ .
\ee
These three equations (\ref{090}), (\ref{121}) and (\ref{721})
will finally fix the flux, dilaton and the warp factor $e^{-2U}$
as well as the geometry of the internal space.  But before we come
to this, we have to discuss the decomposition of the Killing
spinor and the supersymmetry projectors.


\section{Killing spinors and supersymmetric projectors}


Type IIA supergravity has two spinors of opposite chirality and
hence, also the Killing spinor $\epsilon$ should decompose in two
different Majorana-Weyl spinors $\epsilon_{L,R}$ as $\epsilon =
\epsilon_L + \epsilon_R$. In some cases the Killing spinor equations
can be solved with just one Majorana-Weyl spinor ($\epsilon_L$ or
$\epsilon_R$), which simplifies significantly the calculation.  In
general however, this is not the case (as we shall see for the massive
case) and therefore in the decomposition of the 10-d spinor ($\epsilon$)
into 4-d spinors ($\theta$'s) and 6-d spinors ($\eta$'s) one has to
sum over all independent spinors. We shall distinguish the following
two cases:
\be072
\ba{rcl}
(i)  &&  \epsilon = \theta \otimes \eta + \theta^\star \otimes \eta^\star
     \ , \\
(ii) & \quad &  \epsilon = \theta_1 \otimes \eta_1 + \theta_2 \otimes
    \eta_2 + \theta_1^\star \otimes \eta_1^\star + \theta_2^\star \otimes
    \eta_2^\star
\ea
\ee
where $\eta_{\{1,2\}}$ are 6-d chiral spinors and the chirality
properties of the 4-d spinors $\theta_i$ will be fixed later.

\bigskip

\noindent
{\em Comments on case} $(i)$.

This is the Ansatz suitable for massless Type IIA supergravity which
can be lifted on a circle to 11-d supergravity where the internal
space becomes 7-dimensional.  In the simplest situation, there is only
one 7-d Killing spinor $\eta_0$ which can always be written as a $G_2$
singlet.  In the reduction back to 10-d dimensions, the internal space
becomes 6-dimensional, and using $\gamma^7$ we can build two chiral
spinors which can be combined into one complex spinor, representing a
singlet under $SU(3) \subset G_2$ in the following way:
\be138
\eta = {1 \over \sqrt 2} ({\mathbb 1} - \gamma^7) \, e^{\alpha + i \beta}
\, \eta_0
\ee
with $\alpha$ and $\beta$ as real functions.  This is the complex 6-d
spinor $\eta$ appearing in the Ansatz for $\epsilon$ in case $(i)$ and
the function $\alpha$ and $\beta$ have to be fixed by the Killing
spinor equations.  The $G_2$ singlet spinor $\eta_0$ has just one
(real) component and is normalized to $\eta_0^T \eta_0 = 1$.  The
{6-d} $\gamma$-matrices satisfy $(\gamma_m)^T = - \gamma_m =
(\gamma_m)^\star$, which yields for the transposed spinor: $\eta^T =
{1 \over \sqrt{2}} e^{\alpha +i \beta} \eta_0^T ({\mathbb 1} +
\gamma^7)$. Since the internal spinors commute, one obtains the
identities
\be140
0=  \eta^T \eta =
\eta^T \gamma_m \eta = \eta^T \gamma_m \eta^\star =
\eta^T \gamma_{mn} \eta \quad , \qquad \eta^T \eta^\star = e^{2\alpha}\  .
\ee
The complex structure and holomorphic 3-form are introduced as usual
\be150
\eta \, \gamma_{mn} \eta^\star = i \, e^{2 \alpha} \, J_{mn}
\quad , \qquad
\eta \gamma_{mnp} \eta = i \, e^{2 (\alpha + i\beta) }  \, \Omega_{mnp}
\ee
which are related to the $G_2$ invariant 3-form ($\varphi_{rst}
= - i \, \eta_0 \gamma_{rst} \eta_0$) by
\bea232
 J_{mn} &=& \varphi_{mn7} \quad , \\ \label{328}
\Omega_{mnl} &=& \varphi_{mnl} + i \, J_{m}^{\ k} \varphi_{knl}
=\chi^+_{mnl} + i \, \chi^-_{mnl} = i(\delta_m^{\ p} + i J_m^{\ p})
\chi^-_{pnl} \ .
\eea
The properties of the $G_2$-invariant 3-form $\varphi$ yield the
relation $\chi^- = J \chi^+$ which in turn implies that the 3-form
$\Omega$ is holomorphic [$({\mathbb 1} - i\, J)_{mn} \Omega_{npq}
=0$]. Since the spinor $\eta_0$ is a $G_2$ singlet, it has to
satisfying the constraint ($r,s,t, \ldots = 1, \ldots , 7$)
\[
{\mathbb P}_{+\ tu}^{rs}\, \gamma_{rs} \, \eta_0 \equiv
{2 \over 3} \Big( {\mathbb 1}^{rs}_{\ \ tu} +
{1 \over 4} \psi^{rs}_{\ \ tu} \Big)
\gamma_{rs} \, \eta_0 =0
\]
which is the projector onto the ${\bf 14}$ (adjoint of $G_2$) and
$\psi_{pqrs}$ is the $G_2$-invariant 4-index tensor.  This projector
is equivalent to the condition
\be234
 (\gamma_{rs} - i \, \varphi_{rst}\gamma^t) \, \eta_0 =0 \quad ,\qquad
r,s,t = 1 , \ldots , 7 \ .
\ee
We can derive the constraints satisfied by the spinors $\eta$ by
multiplying this equation with $({\mathbb 1} + \gamma^7)$. Using
(\ref{232}) we find
\be230
\ba{rcl}
(\gamma_m  - i J_{mn} \gamma^n ) \, \eta &=& 0 \ , \\
(\gamma_{mn} +  i\, J_{mn}) \, \eta &=& {i \over 2} \,
e^{2i\beta} \Omega_{mnp} \gamma^p  \, \eta^\star \ ,
 \\
(\gamma_{mnp} + 3 i J_{[mn} \gamma_{p]} ) \, \eta &=& i \, e^{2i\beta}
\Omega_{mnp} \eta^\star
\ea
\ee
and employing the projector
\be615
P_\pm \equiv {1 \over 2} ( {\mathbb 1} \pm i J)
\ee
one can decompose the above constraints into holomorphic and
anti-holomorphic components. Using $\{a,b,c\}$ for holomorphic and
$\{\bar a , \bar b , \bar c\}$ for anti-holomorphic indices we
find
\be233
\ba{rcl}
\gamma_a \, \eta^\star = \gamma_{\bar a} \, \eta &=& 0 \ , \\
\gamma_{ab} \, \eta &=&  {i \over 2} \,  e^{2i\beta} \,
\Omega_{abc} \gamma^c  \, \eta^\star \ ,
 \\
(\gamma_{a \bar b} +  i \, J_{a \bar b} ) \, \eta^\star &=& 0 \ .
\ea
\ee
The complex conjugate of these equations gives analogous constraints
for anti-holomorphic indices (note $\gamma^a = \delta^{a\bar b}
\gamma_{\bar b}$). Moreover, one finds
\be237
 \gamma_{abc} \, \eta = i  \, e^{2i\beta} \,
 \Omega_{abc} \, \eta^\star  \ .
\ee
If the spinor $\eta$ is covariantly constant, the six-manifold has
SU(3) holonomy. But non-trivial fluxes will introduce
SU(3)-structures [(con-)torsion] and the space is in general
neither complex, nor K\"ahler nor Ricci flat; we return to this
point in Section V.

\bigskip

\noindent
{\em Comments on case} $(ii)$.

The existence of two (chiral) 6-d spinors $\eta_{\{1 ,2\}}$ implies
the existence of a holomorphic vector ($\sim \eta_1 \gamma_m
\eta_2^\star$) and we can write the two spinors as
\be263
\eta_1 = \eta \quad, \qquad
\eta_2 = v \, \eta \quad , \qquad v = v_m \gamma^m \quad
( v_m v_m = 1 )
\ee
and $\eta$ is given as in eq.\ (\ref{138}). Therefore, the two spinors
have opposite chirality
\be883
\gamma^7 \eta_{1} =  -\eta_{1}
\quad , \qquad
\gamma^7 \eta_{2} =  \eta_{2} \ .
\ee
Note, the complex conjugate spinors $\eta_i^\star$ have opposite
chirality of $\eta_i$.  Since we have not yet specified the spinors
$\theta_i$, we do not make any restriction by the above choice of the
chirality for $\eta_{\{1,2\}}$ (e.g.\ by exchanging $\eta_2
\leftrightarrow \eta_2^\star$ we would have two 6-d spinors of the
same chirality).

The relations for the spinor $\eta$ (\ref{140}), imply now the
following identities (up to the exchange $\eta_1 \leftrightarrow
\eta_2$)
\be612
\ba{l}
\eta_1^T \eta_2 = \eta_1^T \eta_2^\star=
\eta_1^T \gamma^m \eta_2 =\eta_1^T \gamma^m \eta_1^\star = 0 \ , \\
 \eta_1^T \eta_1^\star = \eta_2^T \eta_2^\star =
\eta^T \eta^\star = e^{2\alpha} \ , \\
\eta_1^T \gamma_m \eta_2^\star  =  e^{2\alpha} (\delta_{mn} + i \, J_{mn} )\,
 v^n \ .
\ea
\ee
The last equation implies: $(\delta^{mn} - i \, J^{mn}) \, (\eta_1
\gamma_n\eta_2^\star)= 0$ and therefore this vector is holomorphic.
[Note, in the tangent space there is no distinction between upper and
lower indices ($m,n, \ldots = 1 , \ldots , 6$), which is in contrast
to the holomorphic notation where lowering and rising an index
involves a complex conjugation.]

Using the $\gamma$-identity (\ref{030}), we can derive analogous
projector conditions as in case $(i)$, cp.\  eq.\ (\ref{230}).
Defining
\be227
\hat \Omega_{mn} = \Omega_{mnp} v^p \quad , \qquad
v_\pm^m = {1 \over 2} (\delta^{mn} \pm i J^{mn} ) v_n
\ee
we find
\be782
\ba{rcl}
\gamma^m \, \eta_2 &=& 2 v_-^{m} \, \eta_1
-{i \over 2} e^{2i\beta} \hat \Omega^{mn} \gamma_n \, \eta_1^\star \ , \\
\gamma^{mn}  \, \eta_2 &=& -i J^{mn} v^p \gamma_p \, \eta_1 +
i e^{2i\beta} \hat \Omega^{mn} \, \eta_1^\star -
4 v_-^{[m} \gamma^{n]} \, \eta_1
\ .
\ea
\ee
If one takes into account the holomorphic structure of the quantities,
i.e.\ that $\Omega$ is a (3,0)-form, $\hat \Omega$ a (2,0)-form, $J$ a
(1,1)-form and $v_+$ is a (1,0)-vector, it is straightforward to
express these equations in holomorphic and anti-holomorphic indices
\be762
\ba{lcl}
\gamma_a \, \eta_2 = - {i \over 2} e^{2i\beta} \hat \Omega_{ab} \gamma^b
\, \eta_1^\star
& , &
\gamma_{\bar a} \, \eta_2 = v_{\bar a} \, \eta_1 \ , \\
\gamma_{ab} \eta_2 = i e^{2i \beta} \hat \Omega_{ab} \, \eta_1^\star
& , &
\gamma_{a \bar b} \eta_2 = -i J_{a \bar b} \, \eta_2 +
2 v_{\bar b} \gamma_a \, \eta_1 \ .
\ea
\ee

In the case that both Killing spinors are covariantly constant, also
the vector $v$ is covariantly constant and the holonomy of the
internal space is further reduced to SU(2).  In fact, in this case the
space would factorize into $R^2 \times X_4$, where $X_4$ is a {4-d}
manifold with SU(2) holonomy and the covariantly constant holomorphic
vector identifies the $R^2$ directions. Of course, this is only
possible if the fluxes are trivial and as we will discuss in Section V
the fluxes will deform the internal manifold by non-vanishing torsion
components (SU(2) structures).


\section{Conditions on Fluxes and intersecting D6-brane configurations}


The relevant equations that fix the metric as well as the fluxes
were given by [see eqs.\ (\ref{090} --(\ref{721})]
\bea091
0&=& e^U (\hat \gamma^5 \otimes \partial U ) \, \epsilon
 +{1 \over 4} e^{\phi} \, \Big( m
      [ {\mathbb 1}\otimes {\mathbb 1} ] -
       e^{2U} \hat \gamma^5 \otimes F \gamma^7 \Big)  \epsilon \ , \\
\label{092}
0&=& \Big[ \hat \gamma^5 \otimes
e^U\, (\partial\phi + 3 \partial U) +{\mathbb 1} \otimes
e^{\phi} \Big( m + {1 \over 12} e^{3U}\, H \gamma^7 \Big) \Big] \epsilon \ , \\
\label{093}
0&=&  \Big[ {\mathbb 1} \otimes \Big( \nabla_m +
{1 \over 2} e^{\phi + {U\over 2} } F_{m}  \gamma^7 \Big) -
{1 \over 4}\, \hat \gamma^5 \otimes  e^{{3 \over 2}U} H_m \gamma^7 \Big] \hat
\epsilon
\eea
where $\hat \epsilon = e^{U \over 2} \epsilon$. In solving these
equations, we shall again distinguish the two cases with a single and
two chiral six-dimensional spinors. As we will see, the spinor Ansatz
in case $(i)$ can only be solved for trivial mass parameter and hence
yields the massless case with only D6-branes turned on, whereas the
mass deformation $m$ requires two 6-d spinors as in case $(ii)$.


\subsection{Case $(i)$: Massless case}


Due to the relations (\ref{233}) the different terms become
\[
\ba{rcl}
(\hat \gamma^5 \otimes \partial U)\,  \epsilon &=&
\hat \gamma^5 \theta \otimes \partial^a U  \gamma_a \eta  + cc \ , \\
m\, ( {\mathbb 1} \otimes {\mathbb 1} \, )\epsilon &=&
m\, \theta  \otimes \eta + cc \ , \\
 (\hat \gamma^5 \otimes F \gamma^7) \, \epsilon &=&
- \hat \gamma^5 \theta \otimes ({i \over 2} e^{2i\beta} F^{ab}
\Omega_{abc} \gamma^c
\eta^\star - i F_{a \bar b} J^{a \bar b} \eta) + cc \ .
\ea
\]
These expression have to be inserted in (\ref{091}) and since $\eta$
and $\gamma^a \eta$ are different spinors, we infer
\be695
m = 0 \quad , \qquad F_{a \bar b}J^{a \bar b} = 0
\ee
and
\be214
e^{-U} \partial_a U =  {i \over 8} \, e^\phi \, \Omega_{abc} F^{bc} \ ,
\ee
if the 4-d spinor satisfies the relation
\be211
e^{i\beta } \theta \equiv \hat \theta = \hat \theta^\star \ .
\ee
As we will see below, in order to allow for massive deformations we
need at least two internal spinors. But let us also mention, that a
non-trivial 4-form flux might change the situation, because the 4-form
contribution in the Killing spinor equations can naturally compensate
the mass term (see also the example discussed already by Romans
\cite{110}). We plan to return to issue of non-vanishing 4-form flux
in the future.

Next, consider the eq.\ (\ref{092}) which for $m=H=0$ is trivially
solved by
\be110
\phi = - 3 \, U
\ee
(recall, in our setup $H$ vanishes in massless case).  Finally, we
have to investigate eq.\ (\ref{093}) which becomes for $H=0$
\[
\ba{rcl}
0 &=& \theta \otimes \hat \nabla_m \hat \eta^\star + \theta^\star
\otimes \hat \nabla_m \hat \eta^\star \\
&=& \theta \otimes [\nabla_m - { 1 \over 2} e^{-{5\over 2} U }
F_{mn} \gamma^n ] \hat \eta +
\theta^\star \otimes [\nabla_m + { 1 \over 2}
e^{-{5\over 2} U } F_{mn} \gamma^n ] \hat \eta^\star
\ea
\]
with $\hat \eta = e^{U\over 2}\eta$.  Note, $\eta$ and $\gamma^n \eta$
are spinors of opposite chirality and using (\ref{211}) and collecting
the spinors of the same chirality, we find
\be130
\nabla_m \hat \eta + {1 \over 2} e^{2i\beta} e^{-{5\over 2}U}
F_{mn} \gamma^n \hat \eta^\star = 0 \ .
\ee
If we identify $\alpha = -{U \over 2}$, the spinor $\hat \eta$ is
normalized by $\hat \eta^\star \hat \eta =1$ and thus, multiplying
this equations with $\hat \eta^\star$ and using the relations
(\ref{140}), one finds that $\partial_m \beta = 0$ and this phase can
be dropped in this case. In the next Section, we will use this
differential equation to determine the torsion components.

To summarize, for the spinor Ansatz $(i)$ in (\ref{072}) we found
the following constraints on the fluxes ($m,n, \cdots = 1, \cdots
, 6$)
\be772
\ba{l}
m = H \equiv dB = 0 \quad , \qquad J^{mn} F_{mn} = 0 \ , \\
e^{2U} \partial_q U = - {1 \over 8} \, h_{qp}\, \chi^{-\, pmn} F_{mn}
 \quad , \qquad \phi = - 3U \ .
\ea
\ee
In the special case, that the D6-brane lives in a flat non-compact
10-d space, these results reproduce the known D6-brane solution
given in the string frame by
\be200
\ba{l}
ds^2 = {1 \over \sqrt{H} }\Big[ -dt^2 + d\vec x^2 \Big] + {1 \over \sqrt{H}}
   \Big[ \, dy_1^2 + dy_2^2 + dy_3^2 + H\, (dy_4^2 + dy_5^2 +dy_6^2) \,
     \Big] \\
e^{-4 \phi} = H^3 \quad , \qquad F_{mn} = \epsilon_{mnp} \partial_p H
\ea
\ee
where $H$ is a harmonic function and $e^{2U} = H^{1 \over 2} = e^{- {2
\over 3} \phi}$ and moreover in this case $\chi^-= - dy^4 \wedge dy^5
\wedge dy^6$ \big[or $\chi^{-\, ijk} = - {1 \over \sqrt{g_3}}
\epsilon^{ijk}$, where $g_3 = H^3$ is the determinant of the metric on
the subspace spanned by the coordinates $\{y_4, y_5 , y_6\}$\big].


\subsection{Case $(ii)$: Massive deformation}


As next step we consider in (\ref{072}) the spinor Ansatz $(ii)$ and
start again with the external gravitino variation as given in
eq.\ (\ref{091}).  Using the relations (\ref{233}) [recall $\eta_1 =
\eta$] and (\ref{762}), the different terms become
\be726
\ba{rcl}
(\hat \gamma^5 \otimes \partial U)\,  \epsilon &=&
\hat \gamma^5 \theta_1 \otimes \partial^a U \gamma_a \eta_1
+ \hat \gamma^5 \theta_2 \otimes (\partial^{\bar a} U v_{\bar a} \eta_1
- {i \over 2} e^{2i\beta} \partial^a U \hat \Omega_{ab} \gamma^b \eta_1^\star)
+ cc  \ , \\
m \, ( {\mathbb 1} \otimes {\mathbb 1}) \, \epsilon &=&
 m\, (\theta_1 \otimes \eta_1 + \theta_2 \otimes v^a \gamma_a \eta_1
+ cc )  \ , \\
 (\hat \gamma^5 \otimes F \gamma^7) \, \epsilon &=&
 \hat \gamma^5 \theta_1 \otimes (i F_{a \bar b} J^{a\bar b} \eta_1
-{i \over 2} e^{2i\beta} F^{ab} \Omega_{abc} \gamma^c \eta_1^\star) \\
&& - \hat \gamma^5 \theta_2 \otimes (i F_{a \bar b} J^{a \bar b}
v^c \gamma_c
\eta_1 -i e^{2i\beta} F^{ab} \hat \Omega_{ab} \eta_1^\star + 
2 F^{a \bar b}
v_{\bar b} \gamma_a \eta_1) + cc \ .
\ea
\ee
Now, these expressions have to cancel when inserted into (\ref{091})
and one obtains two complex equations; one proportional to the
spinor $\eta_1$ and the other proportional to $\gamma_a \eta_1$ :
\bea529
0&=& e^{-U} v_{\bar a} \partial^{\bar a} U \, \hat \gamma^5 \theta_2
 + {1 \over 4} e^{\phi} \Big( m \, e^{-2U} \theta_1  -
  i F_{a \bar b} J^{a \bar b} \,  \hat \gamma^5 \theta_1 -
  i  e^{-2i\beta}
  F_{ab} \hat \Omega^{ab} \, \hat \gamma^5 \theta_2^\star \Big) \ ,
\\ \label{242}
0&=& e^{-U} \partial^a U \,\hat \gamma^5 \theta_1 +
{i \over 2} e^{-U} e^{-2i\beta} \partial_b U \hat \Omega^{ba}  \, \hat \gamma^5
\theta^\star_2
\\ \nonumber &&
+ {1 \over 4} e^{\phi} \Big(m v^a\, e^{-2U} \theta_2   - { i \over 2}
e^{-2i\beta} F_{bc} \Omega^{bca}\, \hat \gamma^5 \theta_1^\star
+ i F_{b \bar c} J^{b \bar c} v^a \, \hat \gamma^5 \theta_2
 + 2 \, F^{a \bar b} v_{\bar b} \, \hat \gamma^5 \theta_2
\Big) \ .
\eea
In order to solve these equations, we relate the 4-d spinors
$\theta_1$ and $\theta_2$ by
\be620
\hat \gamma^5 \theta_2 = \theta_1
\ee
and as in case $(i)$ we take again $e^{i \beta} \theta_1 = \hat
\theta$ as a real spinor.  As consequence we get one equation
proportional to $\hat \theta$ and another proportional to $\hat
\gamma^5 \hat \theta$. Since $\hat \theta$ is a Majorana spinor,
these terms have to cancel separately which gives the eqs.
\be412
\ba{rcl}
e^{-U} \partial_a U &=& {1 \over 4} e^{\phi} \Big({i \over 2}
\Omega_{abc} F^{bc} - m\, v_a \, e^{-2U} \Big) \ , \\
i \, e^{-U} \hat \Omega_{ab} \partial^b U & = & - e^{\phi}
F_{a \bar b} v^{\bar b} \ , \\
F_{a \bar b} J^{a \bar b} &=& 0
\ea
\ee
where the (2,0)-form: $\hat \Omega_{ab} \equiv \Omega_{abc} v^c$ was
introduced in (\ref{227}). Inserting the first into the second
equation, we find: $F_{a b} v^b + F_{a \bar b} v^{\bar b} = 0$ and
therefore the 2-form $F$ cannot have components along the vector $v$.

Next, using the same relation as in the massless case: $\phi = -3U$ the
terms in the second equation (\ref{092}) can be written as
\be799
\ba{l}
m\, ( {\mathbb 1} \otimes {\mathbb 1}) \, \epsilon =
m\, (\theta_1 \otimes \eta_1 + \theta_2 \otimes v^a \gamma_a \eta_1
+ cc )  \ , \\
({\mathbb 1} \otimes H \gamma^7) \, \epsilon =
-\theta_1 \otimes  H\eta_1 + \theta_2 \otimes H \eta_2 + cc \ .
\ea
\ee
Because there is no $\hat \gamma^5$, each term proportional to
$\theta_1$ and $\theta_2$ has to vanish separately. Using
(\ref{030}) we find
\[
\ba{l}
H \eta_1 = -3i H^{a \bar b c} J_{a \bar b} \gamma_c \, \eta_1
+ i e^{2i \beta} H^{abc} \Omega_{abc} \, \eta_1^\star \\
\ea
\]
and the term ${\cal O}(\theta_1)$ gives
\[
\Omega_{abc} H^{abc} =12\, i\,  m \quad , \qquad
J_{a\bar b} H^{a \bar b c}= 0 \ .
\]
This relation simplifies also the calculation of $H\eta_2$,
which we write as $H_{npq} \gamma^n \gamma^{pq} \eta_2$ and use
(\ref{782}). As a result we find
\[
H\eta_2 = i e^{2i \beta} \Big( \, {1 \over 2}
H^{abc} \Omega_{bcd} + H_{dbc} \Omega^{bca} \Big)
v_{a} \gamma^d \, \eta_1^\star
\]
and get finally the constraint on the 3-form flux
\[
H^{abc} \Omega_{bcd} = 4 i m \, \delta^a_{\ d} \ .
\]
As the last equation we have to discuss eq.\ (\ref{093}) for the
spinor. Collecting again terms of the same chirality gives now the
following equations for $\hat \eta_{\{1,2\}} \equiv e^{U \over 2}
\eta_{\{1,2\}}$
\bea413
\nabla_m \hat \eta_1 &=& {1 \over 4} e^{{3 \over 2} U} e^{2i\beta}
H_m \hat \eta_2^\star -{1 \over 2} e^{-{5 \over 2} U } e^{2i \beta}
F_m \hat \eta_1^\star\ , \\
\nabla_m \hat \eta_2 &=&  -{1 \over 4} e^{{3 \over 2}U} e^{2i \beta}
H_m \hat \eta_1^\star + {1 \over 2} e^{-{5 \over 2} U }
e^{2i \beta} F_m \eta_2^\star
\eea
and since $\eta_2 = v_m \gamma^m \eta_1$, the second equation fixes
the vector $v_m$.  The first equation on the other hand determines
again the torsion components (see next Section).

To summarize, by solving the Killing spinor equations of massive Type
IIA supergravity (with trivial 4-form flux), we derived the following
conditions for the bosonic background:
\be131
8\, e^{2U} \partial_a U = i\, \Omega_{abc} F^{bc} - 
2 m\, v_a \, e^{-2U}
\ee
and
\be779
F_{ab} v^b + F_{a \bar b} v^{\bar b} = 0 \quad , \quad
F_{a \bar b} J^{a \bar b} = 0 \quad , \quad
H_{a \bar b c} J^{a \bar b} = 0 \quad , \quad
H^{abc} \Omega_{bcd} = 4 i m \, \delta^a_{\ d} \ .
\ee
Recall, the absence of 4-form fluxes implied the constraint: $m (B
\wedge B) = 0$ and bcause $F=m B+dA^{(1)}$ the last condition means:
$^\star(dF \wedge \Omega) \sim m^2$.

An obvious solution is to  keep only the holomorphic components of
the 2-form $F$, i.e.\ to set $F_{a \bar b} =0$.  For the special
case that the D6-branes is embedded into flat space, we find $H
\sim m \chi^-_{mnp} dy^m \wedge dy^n \wedge dy^p = m \, dy^4
\wedge dy^5 \wedge dy^6$ and our results agree with the solution
found in \cite{360} yielding the metric as in (\ref{200}) where
the harmonic function has to be replaced by
\be512
H \rightarrow e^{4U} =m y^1 - \sum_p M_p y^p y^p + H(\vec y)
\quad , \quad \sum M_p = {m^2 \over 2}
\ee
where the vector field is given by $v_m dy^m = dy^1$.  Now, if
$\partial^2 H(\vec y) = -n_6 \delta^3(\vec y) $ this solution
describes $n_6$ (massive) D6-branes and replacing $y^1 \rightarrow
-|y^1|$ corresponds to D8-branes at $y^1 = 0$ (O8-branes
correspond $y^1 \rightarrow |y^1|$, see\cite{290}).

The locations of the different branes can also be identified by
investigating the supersymmetry projectors
\[
\ba{rcl}
\Big( {\mathbb 1} + {1 \over 12 m} H_{MNP} \Gamma^{MNP} \Big) \epsilon &=& 0
\ ,
\\
m\, \Big( {\mathbb 1} + v_M \Gamma^M \Big) \epsilon &=& 0 \ .
\ea
\]
By inserting the $\Gamma$-matrices as given in eq.\ (\ref{070}), the
first equation becomes equivalent to (\ref{092}) [with $\phi = -3U$]
and the second equation is identically fulfilled by our spinor Ansatz
for case $(ii)$ [with (\ref{263}) and (\ref{620})]. In the massless
case, the second projector is empty whereas the first projector gives
the location of the 6-branes, i.e.\ the 3-form defines the 3-d
transversal space of the 6-brane. In the massive case, the second
projector identifies the location of the D8-branes.


\section{Back reaction on the geometry and $G$-structures}


It is obvious that, due to the fluxes, the 7-d spinors are not
covariantly constant and hence also the complex structure $J$ as well
as the holomorphic 3-form $\Omega$ cannot be covariantly constant. The
deviation is related to non-trivial torsion components (or
$G$-structures) and in the following we shall summarize some aspect
relevant for our setup.  For details see e.g.\ \cite{340,170,
Gauntlettetal,Louisetal, LustetalII, LustetalIII,351,650}.  To include
torsion, one replaces the covariant derivative of a spinor $\eta$ by
\[
\nabla_m \eta \rightarrow \big( \nabla_m -
{1 \over 4} \tau_m^{\ pq} \gamma_{pq} \big) \eta
\]
where the 3-index object $\tau$ is the intrinsic torsion\footnote{In
this spinorial context, it is also called (intrinsic) con-torsion.}.
Since the spinor $\eta$ is an SU(3) singlet, $\gamma_{pq} \eta$ does
not contain the adjoint of SU(3) and thus the intrinsic torsion is an
element of $\Lambda_1 \otimes su(3)^\perp$, where $\Lambda_1$ denotes
the space of 1-forms and $su(3)^\perp$ denotes the compliment to the
SU(3) Lie algebra, i.e.\ $su(3) \oplus su(3)^\perp = so(6)$.  Thus,
although the 3-index object $\tau$ can have ${\bf6} \otimes {\bf 15}$
components, only $ {\bf 6} \otimes {\bf 7}$ components contribute to
the intrinsic torsion and these components are decompose under SU(3)
as follows
\[
{\bf 6} \otimes {\bf 7}
\rightarrow \bf (1 + 1) \oplus (8 + 8) \oplus (6 + \bar 6)
\oplus (3 + \bar 3) \oplus (3 + \bar 3)
 = {\cal W}_1 \oplus {\cal W}_2 \oplus {\cal W}_3
\oplus {\cal W}_4 \oplus {\cal W}_5
\]
where ${\cal W}_1$ is a complex scalar, ${\cal W}_2$ a 2-form, ${\cal
W}_3$ a 3-form and ${\cal W}_4$ as well as ${\cal W}_5$ are two
vectors. These components can now be read-off from $dJ$ and $d\Omega$
as
\be524
\ba{rcl}
dJ &=& {3 i \over 4}\,  ( {\cal W}_1 \bar \Omega -
 \bar{\cal W}_1 \Omega ) +  {\cal W}_3 + J \wedge  {\cal W}_4 \ ,\\
d\Omega &=&   {\cal W}_1 J \wedge J + J \wedge  {\cal W}_2
+ \Omega \wedge  {\cal W}_5
\ea
\ee
with the constraints
\[
J \wedge J \wedge {\cal W}_2
=J \wedge  {\cal W}_3 = \Omega \wedge {\cal W}_3=0
\]
and therefore ${\cal W}_2$ and ${\cal W}_3$ are a primitive two- and
and three-form, respectively. By using the definition of $J$ and
$\Omega$ in terms of the spinor $\eta$ (see eq.\ (\ref{150}) and
applying Fierz re-arrangements, one can also verify the usual
relations
\[
J \wedge J \wedge J = {3 i \over 4} \, \Omega \wedge \bar \Omega
\quad , \qquad J \wedge \Omega = 0 \ .
\]
The components of ${\cal W}_{\{1,4,5\}}$ can also be written as
\be321
{\cal W}_1 \sim \Omega^{pmn} \partial_p J_{mn}\ ,\quad
({\cal W}_4)_p \sim J^{mn} \partial_{[p} J_{mn]}\ , \quad
({\cal W}_5)_p \sim (\Omega^\star)^{mnq} \partial_{[p} \Omega_{mnq]}\ .
\ee
Depending on the components which are non-trivial, one distinguishes
between different complex and non-complex manifolds. For example, the
manifold is non-complex if $\tau \in {\cal W}_1$ (nearly K\"ahler) and
$\tau \in {\cal W}_2$ (almost K\"ahler) and examples of complex
manifolds are $\tau \in {\cal W}_3$ (special-hermitian), $\tau \in
{\cal W}_5$ (K\"ahler) and of course if $\tau=0$ we have a Calabi-Yau
space (see \cite{630, 340, LustetalII} for more examples).  Let us now
determine the different components for our flux compactification.

\bigskip

\noindent
{\em Massless case $(i)$}

\noindent
For this case $dF=0$ and the spinor has to satisfy the equation
\[
\nabla_m \hat \eta + {1 \over 2} e^{-{5\over 2}U}
F_{mn} \gamma^n \hat \eta^\star = 0
\]
where we set $\beta = 0$ and $\alpha = - {U \over 2}$ so that $\hat
\eta = {1 \over \sqrt{2}} (\1 - \gamma^7) \eta_0$ which is normalized
as $\hat \eta^T \hat \eta =1$ [see (\ref{131}) and (\ref{138})].
Using the fact, that $\hat \eta$ is a SU(3) singlet so that the
conditions (\ref{230}) are satisfied, we can solve this equation by
writing the covariant derivative as
\[
\nabla_m \hat \eta = \partial_m \hat \eta + { 1\over 4} \omega_m^{pq}
\gamma_{pq} \hat \eta = {1 \over 4} \, \omega_m^{pq} \, \Big[ - i
J_{pq} +{ i \over 2} \Omega_{pqr} \gamma^r
\Big] \hat \eta
\]
and since $\hat \eta=const.$ we get first order differential equations
for the Vielbein $e_r^{\ s}$: $0=\omega_m^{pq} \, J_{pq} $
and $\omega_m^{pq} \Omega_{pqs} e^{\ s}_r \sim F_{mr}$.

In order to get the torsion components, we will consider the complex
structure $J$ as well as the holomorphic 3-form $\Omega$ written
in terms of the spinor $\hat \eta$ and find for the covariant
derivative
\bea672
D_p J_{mn} &=& -e^{-{5 \over 2} \, U}  F_{p}^{\ r} \, \chi^-_{rnl} \, J_{lm}
= {1 \over 2} e^{-{5 \over 2} \, U}  F_{p}^{\ r} \, (\Omega_{rmn}
+ \Omega^\star_{rmn})\ , \\
D_p \Omega_{mnq}  &=&   {i\over 2} \, e^{-{5 \over 2} \, U} \,
F_{p}^{\ r} \, J_{r[m} J_{np]}
\eea
where $m,n = 1,\ldots,6$. Using the formulae (\ref{321}) we find
\[
{\cal W}_1 = 0 \quad , \qquad ({\cal W}_4)_m \sim ({\cal W}_5)_m
\sim \chi^-_{mpq} F^{pq} = -8 e^{2U} \partial_m U
\]
where we used in the last equation the monopole equation (\ref{772}).
So, the non-zero values of ${\cal W}_{\{4,5\}}$ are related to the
non-trivial warping of the metric.  In order to fix the remaining
components, we should use holomorphic coordinates and we can write
$d\Omega \sim F \wedge J$.  Since the (3,0) part in $d\Omega$ vanishes
(${\cal W}_1 = 0$) we infer that
\be322
{\cal W}_2 \sim F^{(1,1)}
\ee
On the other hand, since ${\cal W}_1$ vanishes $dJ$ has only a (2,1)
and (1,2) part and therefore only the $F^{(2,0)}$ part and its complex
conjugate contributes to $dJ$.  But since this holomorphic part of $F$
is equivalent to $\Omega_{abc} F^{bc} \sim ({\cal W}_{\{4,5\}})_a$ we
conclude
\be119
{\cal W}_3 = 0 \ .
\ee
These results are in agreement with those derived in \cite{351}.

\bigskip

\noindent
{\em Massive case $(ii)$}

\noindent
It is now straightforward to repeat the analysis for the massive case,
where $dF = mH$ and the differential equation for the spinor $\eta_1$,
which defines $J$ and $\Omega$, was given in (\ref{413}) and can be
written as
\be792
\ba{rcl}
\nabla_m \hat \eta
&= & M_m \hat \eta  + e^{2i \beta} \, N_{mn} \gamma^n \hat \eta^\star \ , \\
{\rm with:} \quad
M_m & \equiv & - {i \over 4} e^{{3 \over 2} U }H_{mpq}
(\hat \Omega^{\star})^{pq} \ , \\
N_{mn} & \equiv &- {1 \over 2} \, e^{-{5 \over 2} U }
\Big(\,  F_{mn} - 2 e^{4U} H_{mnp} v^p_+  \, \Big) \ .
\ea
\ee
Again, $\hat \eta = e^{U \over 2} \eta_1$ and we identified again
$\alpha = - {U \over 2}$ and introduced $M_m$ and $N_{mn}$ to simplify
the notation (note $\beta$ is non-trivial in this case).  One gets
again a set of first order differential equations for the spin
connection, if one uses the fact that $\hat \eta$ is an SU(3) singlet,
i.e.\ obeys the relations (\ref{230}).  If one further takes into
account the non-trivial phase $\beta$, this calculations is analogous
to the massless case.

The covariant derivatives of $J$ and $\Omega$ now become
\[
\ba{rcl}
D_p J_{mn} &=& 2 M_p J_{mn} - N_{p}^{\ q}
\big(\,\Omega^\star_{qmn} + \,\Omega_{qmn} \, \big)
\ , \\
D_q \Omega_{mnp} &=& 2 (M_q -i \, \partial_q \beta)\, \Omega_{mnp}
+ { 2 i} \, N_q^{\ r} \, J_{r[p} J_{mn]}
\ea
\]
where $m,n, \cdots = 1 , \cdots, 6$.  Using holomorphic coordinates,
we find again
\[
\ba{rcl}
{\cal W}_1 &= &0 \quad , \\
({\cal W}_4)_a & \sim &  4 M_a + i \, \Omega_{abc} N^{bc} \ ,\\
({\cal W}_5)_a & \sim & {3} (M_a - i \partial_a \beta )
+ {i} \, \Omega_{abc} N^{bc}
\ea
\]
where we used now the massive monopole equation (\ref{412}) or
(\ref{131}), combined with (\ref{779}).  To find ${\cal W}_2$, it is
enough to look on the last term in $d \Omega$, which is proportional
to $J \wedge N$ and hence
\[
({\cal W}_2)_{a \bar b}  \ \sim\
 N_{a \bar b} \ \sim \ F_{a \bar b} -{2} e^{4U} \,
H_{a \bar b c} v^c \ .
\]
which is primitive because $H$ and $F$ are primitive.  Finally, to get
${\cal W}_3$, we have to consider the (1,2)-piece of $dJ$ which is not
part of ${\cal W}_4$. Therefore, only the term $N_{ab} \Omega^b_{\
\bar c \bar d}$ and its complex conjugate can contribute to ${\cal
W}_3$. However, since $N_{ab} \sim \Omega_{abc} \partial^{c}U$ (i.e.\
$\partial_c U \sim \Omega_{cab} N^{ab}$ which follows from (\ref{131})
and (\ref{779})) and using the identity $\Omega_{abc} \Omega^{cde}
\sim \delta_a^{\ [d} \delta_b^{\ e]}$ we find that all terms of $dJ$
are part of ${\cal W}_4$ and conclude that also for the massive case
\[
{\cal W}_3 =0 \ .
\]
The appearance of the vector $v$ implies a breaking of the SU(3) to
SU(2) structures. In order to decompose our expressions in SU(2)
representations we have to separate the components of ${\cal W}_2$,
${\cal W}_4$ and ${\cal W}_5$ parallel and transverse to $v$. The
(1,1)-form ${\cal W}_2$ decomposes into: ${\bf 1+ (2 + \bar 2) + 3}$
given by $v^a N_{a \bar b} v^{\bar b}$, $N_{a \bar b} v^{\bar b}$,
$N_{\bar a b }v^b$ and the remaining components comprise the ${\bf
3}$. Similarly, by contracting the vectors ${\cal W}_{\{4,5\}}$ with $v$,
we get a ${\bf 1}$ and the remaining components become ${\bf 2 + \bar
2}$.


\section{Interesecting branes and chirality}


In the limit of vanishing mass parameter, our results are
invariant under SU(3) rotations and therefore intersecting brane
solutions can be build by SU(3) rotations as proposed in
\cite{130}. A non-zero mass parameter implies a massive NS-NS
$B$-field yielding a 3-form flux ($H=dB$) and as we discussed this
mass parameter can only be non-zero, if the 6-manifold allows for
a (no-where vanishing) vector $v$. This puts already constraints
on the (compact) manifold, as e.g.,\ a vanishing Euler number
(Hopf theorem), and corresponds to the existence of  two 6-d
spinors with opposite chirality. The massless case on the other
hand, is described by a single 6-d spinor, which is a SU(3)
singlet.

{From} the supergravity point of view, a mass parameter is related to
the appearance of D8-branes which are perpendicular to the vector $v$.
At the same time, this vector breaks the SU(3) rotations known from
the massless case to SU(2) rotations and therefore the D6-branes can
be localized only in four of the six internal directions and are
aligned along one internal direction. An example is given by the
following picture
\renewcommand{\arraystretch}{1}
\[
\begin{array}{c|cc|cc|cc}
   & y_1 & y_2 & y_3 & y_4 & y_5 & y_6  \\ \hline
D6 & \times  &  o   & \times & o &  \times   &  o    \\
D6'& \times  &  o   &   o     &    \times    & o  &\times \\
D8 &  & \times & \times & \times & \times & \times   \\
\end{array}
\]
where $y_1 , \ldots , y_6$ comprises the internal directions and
``$\times$'' indicates the world-volume directions of D6-branes
and ``$o$'' the constant $H$-flux, e.g.,\ in the simplest case: $H
= \, (h\, dy^4 \wedge dy^6 + h'\, dy^3 \wedge dy^5)\wedge dy^2$
for some constants $h, h' \sim m$. Due to the constraint $B \wedge
B = 0$ (coming from $F_4 = 0$), $F$ becomes
\be621
B = (h \, y^4 dy^6 + h'\, y^3 dy^5) \wedge dy^2 \ .
\ee
{From} the supergravity point of view one should distinguish
between localized branes and branes that are dissolved into fluxes
and so far we discussed only the latter ones.  However,  it is
straightforward to add also localized branes. E.g.,\ we can add
$n_6$ localized D6-branes by changing the Bianchi identity
\be520
dF = m H \rightarrow m H - n_6 \, \delta^{(3)}
\ee
where $\delta^{(3)}$ is a 3-form $\delta$-function which projects
onto the world-volume of the D6-brane. In order to add sources for
D8-branes we replace $m \rightarrow - m\, \theta(y_1 - y^0)$ and
NS5-brane source correspond to $dH = n_5\, \delta^{(4)}$, where
the 4-d $\delta$-function projects onto the NS5-brane world-volume
and $n_5$ is the number of NS5-branes. If we ignore for the moment
D8-brane sources and consider only NS5- and D6-branes, only, then
from $ddF =0$ one infers that the D6-branes {\em must} end
on the NS5-branes and the number of D6-branes is given by the
number of NS5-branes and the mass parameter \cite{HZ}:
\[
ddF = 0 = (m n_5 - n_6) \, \delta^{(4)} \ .
\]
Therefore, if one adds localized NS5-branes, one has necessarily
to include open D6-branes that end on these NS5-branes.

With the intersecting D6-brane configuration discussed above we would
now like to address whether the D6-brane world-volume 4-d gauge theory
is chiral. This can be addressed by studying a possible anomaly inflow
\cite{220} from the bulk to the 4-d sub-space. [Without invoking the
constraints imposed by supersymmetry, this anomaly inflow in the
presence of a NS-NS 3-form flux has also been discussed in
\cite{190}.]  Anomaly inflow \cite{220} takes place when the
Wess-Zumino action associated with the given Dp-brane is not invariant
under the gauge transformation. In this case the anomaly of the
world-volume field theory has to cancel the anomaly inflow
contribution, thus rendering the gauge theory chiral. The Wess-Zumino
action associated with the specific Dp-brane world-volume has the
form:
\begin{equation}
 S_{WZ}^p = \int_{Dp}{\cal C} \wedge Y \label{WS}
 \ee
where ${\cal C}$ is a sum over all R-R potentials, $Y \equiv ch(F)
\sqrt{\hat A(R)}$ with $ch(F)$ denoting the Chern class of the
world-volume gauge bundle and $\hat A$ is the A-roof genus, which
depends on the curvature form (see \cite{220} and references
therein). In our specific consideration $X_6$ is flat and $\hat A$
plays no role.

Since $Y$ is exact, it can be written as $Y = dY^{(0)}$ and we can
integrate (\ref{WS}) {\it per partes} to obtain an integral $\int_{Dp}
d{\cal C} \wedge Y^{(0)}$.  Now, $Y^{(0)}$ transforms under a gauge
transformation ($\delta Y^{(0)} = d Y^{(1)}$) and if we denote the
field strengths of the R-R gauge potentials by ${\cal G} \equiv
d{\cal C}$, one finds for the variation of the Wess-Zumino action
(\ref{WS}):
\begin{equation}
 \delta S^p_{WZ} = \int_{Dp} d {\cal G} \, Y^{(1)} \ .
\label{WZv}
\ee
Whenever $d{\cal G}$, projected on the Dp-brane world-volume, is
non-zero, (\ref{WZv}) is non-zero and thus Wess-Zumino action
(\ref{WS}) is not gauge invariant. Its contribution should then be
cancelled by the gauge-anomaly contribution of the D-brane
world-volume gauge theory.

We shall now apply  this inflow mechanism for the world-volume
theory of a D6-brane, i.e.\ $d{\cal G} = dF$. As seen from
eq.\ (\ref{520})  both the NS-NS 3-flux as well as D6-brane
sources can potentially contribute to the anomaly inflow. 
The integral (\ref{WZv}) is non-zero only if the $dF$
projected onto the world-volume of one D6-brane is non-zero.
However, the constraints on the configuration are such that this is
not the case. As it is obvious from the example in the table
above, the $H$-flux always extends in the $y^2$ direction, which is
not part of any of the D6-brane world-volume directions. The same
is also true for the D6'-source term  when we consider the
Wess-Zumino term for D6-brane (and vice versa): $\delta^{(3)}$
includes always the $\delta$-function in $y^2$ direction. Hence
neither the NS-NS 3-form fluxes (or NS5-brane sources) nor the
D6-brane  sources  can  give rise to a non-zero anomaly inflow
term.

Note however, that in our analysis we have to include also effects
from D8-branes, which appear as  domain walls in the common
world-volume direction of D6-branes. If we put them at $y_1=0$ we have
to replace the two-form field strength $F$ with $ - m\,
\theta(y_1) \, B +dA_1$ and the effect of D8-brane sources modifies the
right hand side of the Bianchi identity for $F$ in the following
way:
\begin{equation} dF = - \delta(y_1) \, v \wedge B - m \, \theta(y_1) \,
H \ . \label{mbi} \ee
Recall, the vector field $v$ is orthogonal the D8-branes. Again, the
$H$-term cannot give a non-zero contribution to the inflow
integral. In addition also the first term, which describes the
coupling of the D8-brane background to the world-volume of a D6-brane,
can only be nonzero if the $B$-field, projected onto the D6-brane
world-volume, is nonzero. This however, is only possible if $B \wedge
B \neq 0$ (because the $B$-field has always components in the
transverse space of the D6-brane). However, for our configuration,
without R-R 4-form fluxes, $B\wedge B=0$, and thus there is no anomaly
inflow from D8-brane sources.

The same analysis could be repeated for the Wess-Zumino coupling
for the world-volume of the D8-brane, again  giving no anomaly
inflow. Hence, we conclude that for our specific supersymmetric
configuration (without R-R 4-form turned on), there is no anomaly
inflow from the bulk to the world-volume field theory  and thus the
gauge theory is not chiral.


\section{Discusssion}


We have discussed the constraints imposed by supersymmetry of D6-brane
configurations in massive type IIA superstring theory. In the simplest
case of parallel 6-branes wrapping a 3-cycle of a torus, the massive
deformation of the warp factor or dilaton is given in eq.\ (\ref{512})
with the metric (\ref{200}), which agrees with the result found in
\cite{360}. The supergravity solution exhibits a naked singularity at
a finite distance in the transversal space, which is given by a zero
of $e^{4U}$, which is reminiscent to the deformation of the M2-brane
due to a self-dual 4-form potential \cite{280}.  A similar singularity
occurs in brane world scenarios with positive tension branes where the
warp factor has a zero and in the AdS/CFT language, this singularity
was resolved by non-trivial IR effects. [Note, from the world-volume
point of view the IR regime corresponds to a small warp factor.]  In
the case at hand however, the 10-d warp factor in both, the Einstein-
and string frame, is infinite at the singularity indicating that the
theory is UV ``incomplete''.  Better understanding of this singularity
deserves further investigations. Note, however, that in the case of
non-flat internal space $X_6$ one may allow for the non-singular
configurations with the NS-NS 3-form flux corresponding to a regular
$L^2$ integrable harmonic 3-form on $X_6$, which would in turn, due to
transgression Chern-Simons terms, render the R-R 2-form field strength
and the metric regular, thus making the configuration regular (c.f.,
\cite{CLP}).

\bigskip

\noindent
{\em The superpotential and fixing of moduli}

\medskip

\noindent We would also like to comment on  moduli dependence of
the 4-d superpotential $W$ generated by fluxes.  A non-zero
superpotential  can be determined from the Killing spinor equation
for  a 4-d spinor  $\theta$ which is not covariantly constant, but
satisfied  the relation, written schematically as $D_\mu \theta
\sim W \gamma_\mu \theta$. Implementing this relation into the
calculation of the Killing spinor equations  we obtain the
following two contributions to the superpotential $W$:
\[
\sim \int F \wedge J \wedge J
\quad , \qquad
\sim 12  m \, i +  \int H \wedge \Omega
\]
For our vacuum the superpotential $W$ and its K\"ahler covariant
derivatives vanish.  A contribution  to $W$ from the 2-form $F$
flux yields a dependence of the  K\"ahler class moduli whereas
that from  $H$ yields a dependence on the complex structure
moduli. Note, however, that the 2-form $F$ flux has to satisfy the
constraint $F \wedge F = 0$ and it has to be transverse to the
vector $v$ ($F_{mn} v^n=0$) and therefore it cannot fix the
2-cycle which is related to the holomorphic vector $v$. In
addition, since $dF =mH$, the constraints on $F$ imply analogous
constraints on the 3-form $H$ and hence the contribution to $W$
from $H$  yields a dependence on only some complex structure
moduli.

A way to understand the moduli dependence of the superpotential and
the fixing of moduli in the vacuum is to consider the supergravity
theory obtained after dimensional reduction (see \cite{PS,260,240}).
One obtains moduli from K\"ahler class deformations which are in Type
IIA string theory related to scalar fields in vector multiplets and
complex structure moduli related to scalar fields in hypermultiplets.
General R-R 2- and 4-form fluxes yield a (complex) superpotential that
fixes in the generic case all scalars in the vector multiplets and the
vacuum is described by a BPS domain wall solution of N=2 supergravtiy
\cite{250} which becomes flat space time in the limit of vanishing
superpotential. An additional NS-NS 3-form flux will result in an
additional dependence of $W$ on the complex structure moduli. Our case
at hand, however, is not generic because we have only special R-R
2-form and NS-NS 3-form fluxes which are subject to the constraints
mentioned above. The absence of a R-R 4-form flux would also result in
the absence of certain K\"ahler class moduli in the
superpotiential. (Due to the presence of the NS-NS 3-form flux we
cannot ``turn on'' the 4-flux by a symplectic rotation; see also
\cite{240}.)  In conclusion, due to the constrained structure of the
turned on fluxes only a limited class of moduli can be fixed in the
vacuum.

\bigskip

\noindent
{\em Further open questions}

\medskip

\noindent {There are a number of directions for further
exploration.  The construction described in this  paper  was
severely constrained due to the existence of the vector field $v$,
which, e.g.,\ has forced us to align all D6-brane along this
vector in the internal space. There is however a possibility of a
more  more general setting that does not require the existence of
such a vector. This seems to be possible if one allows for
additional R-R 4-form fluxes. In this case the  additional terms
in the Killing spinor equations can in fact naturally compensate
for  the mass terms without the necessity of two (opposite
chirality)  6-d internal spinors. There is a strong indication
that the inclusion of a R-R 4-form fluxes (and D4-brane sources)
would yield a  chiral gauge theory  on the D4-brane world-volume
(as considered  in \cite{HZ} in the absence of supergavity
fluxes). One also expects that in this case most of  the moduli
(except the dilaton) would be fixed in the N=1 supersymmetric
vacuum.

An alternative approach to derive 4-dimensional N=1
supersymmetric solutions of (massless) Type IIA superstring theory
with intersecting D6-branes and supergravity fluxes would be to
address its strongly coupled limit as M-theory compactified on
7-dimensional manifold with 4-form field strength G fluxes
turned on, resulting in manifolds with $G_2$ structures with
torsion.  Reduction of this seven-dimensional space on a circle
would in turn yield  a (massless) Type IIA superstring theory with
intersecting D6-branes (and O6-planes) and a six-dimensional
manifold with SU(3) torsion classes. This approach may shed a
complementary light on the possible intersecting D6-brane
configurations and the  spectrum of the resulting $N=1$
supersymemtric  D6-brane world-volume gauge theory in the presence
of fluxes and is a subject of further study. Some work in this
direction has been done already, e.g., in \cite{640,140,620}.
However,  a more  general study of the possible flux configurations
and  the implications for the D6-brane world-volume gauge theory
is a subject of further research.

\section*{Acknowledgments}

We would like to thank Ralph Blumenhagen, Amihay Hanany, Dario
Martelli and Angel Uranga for useful discussions. M.C. would like to
thank Institute for Advanced Study, Princeton, and we would like to
thank CERN, Theory Division, for hospitality and support during the
course of this work.  Research is supported in part by DOE grant
DOE-EY-76-02-3071 (M.C.), NSF grant INT02-03585 (M.C.) and Fay R. and
Eugene L. Langberg endowed Chair (M.C.).



\providecommand{\href}[2]{#2}\begingroup\raggedright\endgroup


\end{document}